\documentclass[12pt]{article}

\newfont{\tenmsb}{msbm10 scaled\magstep1}
   
\makeatletter
\@addtoreset{equation}{section}
   
\makeatother

\newcommand\half{{\scriptstyle{\frac{1}{2}}}}
\newcommand{\IR}{{\bf R}}
\newcommand{\vx}{{\vec x}}
\newcommand{\vv}{{\vec v}}
\newcommand{\vnabla}{{\vec\nabla}}
\newcommand{\vp}{{\vec p}}
\newcommand{\vA}{{\vec A}}

\begin{document}

\title{Non-commuting coordinates in
vortex dynamics and in the Hall effect\\
related to ``exotic'' Galilean symmetry}

\author{P.~A.~Horv\'athy
\\
Laboratoire de Math\'ematiques et de Physique Th\'eorique\\
Universit\'e de Tours\\
Parc de Grandmont\\
F-37 200 TOURS (France)\\ 
E-mail: horvathy@univ-tours.fr}  

\maketitle

\begin{abstract}
    Vortex dynamics in a thin superfluid ${}^4$He film
    as well as in a type II superconductor
    is described by the classical counterpart of
    the model advocated by Peierls, and used for deriving
    the ground states of the Fractional Quantum Hall Effect.
    The model has non-commuting coordinates, and is obtained
    by reduction from a particle associated with the ``exotic''
     extension of the planar Galilei group.
\end{abstract}
    
\section{Vortex dynamics and the Peierls substitution}
                                             
(Quantum) Mechanics with non-commuting coordinates\cite{NC,NC2},
\begin{equation}
\big\{x,y\big\}=\theta,
\label{xycomrel}
\end{equation}
has become the focus of recent research.
Such a relation may appear rather puzzling. Below we argue,
however, that it is inherent in a number of physical instances,
and could indeed have been recognized many years ago.

Our first example of non-commuting coordinates is provided by
the effective dynamics of point-like flux lines in a thin film of
superfluid ${}^4$He \cite{Onsager,MaWe}.
For the sake of simplicity, we restrict ourselves to two 
vortices of identical vorticity. 
The center-of-vorticity coordinates are constants of the motion. 
For the relative coordinates 
$x=x_{1}-x_{2}$ and $y=y_{1}-y_{2}$, respectively,
the equations of motion become\cite{Onsager,MaWe,HMC}
\begin{eqnarray}
    (\rho L\kappa)\,\dot{x}=\partial_{y} H,
    \qquad
    (\rho L\kappa)\,\dot{y}=-\partial_{x} H,
     \label{effdyn}
\end{eqnarray}
where $\rho$ and $ L$ are the density and the thickness of the 
film, respectively; $\kappa$ is the (quantized) vorticity.  
The Hamiltonian reads 
\begin{equation}  
H=-\frac{\rho L\kappa^2}{4\pi}\ln r.
\label{vortham}
\end{equation}
\goodbreak

Eq. (\ref{effdyn}) is plainly a Hamiltonian system, 
\begin{equation}
	\dot{\xi}=\big\{\xi,H\big\},
	\qquad
\xi=(x,y),
\label{hameq}
\end{equation}
with the Poisson bracket associated with the symplectic structure 
of the plane,
$ 
\theta^{-1}\,dx\wedge dy,
\,
\theta\equiv(\rho L\kappa)^{-1}.
$
Vortex dynamics in a superfluid helium film provides us therefore
with non-commuting coordinates, since (\ref{xycomrel}) holds.
Let us emphasize that the symplectic plane should be viewed as
the classical {\it phase space}.

The classical motions are determined at once: owing to energy conservation
and consistently 
with the conservation of the vorticity, the motions 
are uniform rotations with angular velocity is
$-(\kappa/2\pi)r^{-2}$. 

The equations (\ref{hameq}) can be derived\cite{MaWe} from the 
Euler equations of an incompressible fluid, 
$\dot{\vv}+\vv\cdot\vnabla\vv=-\vnabla p$,
$
\vnabla\cdot\vv=0,
$
where $p$ is the pression. Defining the {\it vorticity field} 
as $\omega=\vnabla\times\vv$ and taking the curl of 
the Euler equations yields 
\begin{equation}
    \dot{\omega}+\vv\cdot\vnabla\omega=0.
\label{vorteq}
\end{equation}

These equations are Hamiltonian,
$\dot{\omega}=\big\{\omega,H\big\}$, with 
\begin{equation}
    \big\{F,G\big\}=\int\omega\left\{\frac{\delta F}{\delta\omega},
    \frac{\delta G}{\delta\omega}\right\}dxdy
    \quad\hbox{and}\quad
    H=\int\!
    \half L\rho\vv^2dxdy,
\label{Hamstruct}
\end{equation}
where $\big\{\cdot,\cdot\big\}$ is the 
Poisson bracket on the symplectic plane. By incompressibility,
$\vv=\vnabla\times\psi$, so that $H=-(L\rho/2)\int\omega\psi dxdy$.
Then, assuming that the vorticity is supported by pointlike objects, 
$\omega=\sum_{i}\kappa\delta(\vx-\vx_{i})$, 
eqns. (\ref{vortham}) and (\ref{hameq}) follow.

Vortices in a thin superconducting film 
behave in a similar manner\cite{Fetter}.
Here one starts with the  magnetohydrodynamic equations
$
\dot{v}_{i}-\epsilon_{ij}v_{j}\vnabla\times\vv
=(e/m)\big(E_{i}+\epsilon_{ij}v_{j}B\big)
$
of  a charged incompressible fluid.
The vorticity, which reads now rather $\omega=\vnabla\times\vv+eB/m$, 
satisfies the same equation, Eq. (\ref{vorteq}), as in the superfluid 
case. Then the generalized London equation  requires
\begin{equation}
    \vnabla\times\vv+eB/m=(h/2e)\sum_{i}\delta\big(\vx-\vx_{i}\big)
    \label{London}
\end{equation}
where $h/2e$ is the flux quantum. The vorticity is hence supported
again by pointlike objects, and
one recovers the same model as for a neutral superfluid.
The Hamiltonian, $H$, is the sum of the interaction 
potentials\cite{Fetter}. 

Yet another example is provided by
a  extreme type II bulk superconductor\cite{Fetter}. 
The mean distance of the quantized flux lines
is much larger than their core size, so they can be viewed as
 vortex filaments in an incompressible and frictionless fluid.
 Using the London equation it is shown\cite{Fetter} that
 the displacements of the vortices from their
 fixed positions satisfy, to first order, 
 the same pairs of equations above (\ref{effdyn}), i. e.
(\ref{hameq}). The Hamiltonian is the sum of more involved
non-local expressions\cite{Fetter}. 
\goodbreak

 Another peculiarity is the absence of a mass term in the hamiltonian
(\ref{vortham}). 
The analogy with the motion of {\it massless} particles in a
magnetic field, noticed before\cite{HMC},
can be further amplified.
The model can indeed also obtained by considering the
massless limit of an ordinary, charged particle in the plane
subject to an electromagnetic field\cite{DJT}.
The $m\to0$ limit yields in fact the first-order Lagrangian
without mass term,
\begin{equation}
	L=\frac{eB}{2}\big(x\dot{y}-y\dot{x}\big)-eV,
\label{redlag}
\end{equation}
whose associated Hamiltonian system is (\ref{hameq}) with 
$\theta^{-1}=eB$ and $H=eV$.

Quantizing the classical system (\ref{hameq}) yields the so-called
he ``Peierls substitution''.
Seventy years ago Peierls\cite{Peierls} argued in fact that
in a strong  magnetic field $B$ and weak potential $V$
the electrons remain in the lowest Landau 
level, and the energy is simply  
$
E_{n}={eB}/{2m}+\epsilon_{n},
$
where the
$\epsilon_{n}$ are the eigenvalues of the operator 
$\hat{V}(\hat{x},\hat{y})$, obtained from the 
potential alone, but such that $\hat{x}$ and $\hat{y}$ are
canonically conjugate,
\begin{equation} 
[\hat{x},\hat{y}]=\frac{i}{eB}. 
\label{NCcoord}
\end{equation}

In more recent times, the ``Peierls substitution'' has re-emerged 
in the theory of the Quantum Hall Effect\cite{LAUGH,GJ,QHE}.
It is argued\cite{LAUGH} that the ``$\frac{1}{3}$'', or more generally,
the fractional effect arises from  
``condensation into a collective ground state which
 represents a novel state of matter''\cite{LAUGH}.
 This latter
consists of a multi-electron system represented  by Laughlin's
 quasiparticles\cite{QHE}, all of which lie  in the lowest Landau
 level and obey hence the Peierls dynamics. 

 The clue of the relation between the fractional Hall effect and
superfluid vortices is that Laughlin's quasiparticles correspond to
 the vortex solutions in the effective Landau-Ginzburg field theory
 of the QHE\cite{ZHK}. The simple hamiltonian system 
(\ref{hameq}) lies hence at the heart of this deep relation.
\goodbreak

\section{Exotic particles and the Hall effect}

A slightly different derivation\cite{DH} of the system (\ref{hameq})
starts with  a particle associated with the ``exotic'' two-parameter 
central extension of the planar Galilei group \cite{exotic}.
Let us describe this point in some detail.

According to geometric quantization\cite{SSD,GQ}, elementary particles
correspond to coadjoint orbits of their fundamental symmetry groups,
endowed with their canonical symplectic structures. These latter are 
labeled in turn by the cohomology classes. In any dimension $d\geq3$,
the Galilei group has one-dimensional cohomology labeled by a
real parameter $m$, identified with the mass. 
The {\it planar} Galilei group admits however, a second cohomology 
label, $k$ \cite{exotic}. The orbits are still $\IR^4$, but the 
corresponding ``exotic'' symplectic structure is, rather,
\begin{equation}
    d\vp\wedge d\vx+\frac{\theta}{2}\epsilon_{ij}dp_{i}\wedge dp_{j},
    \qquad
    \theta\equiv\frac{k}{m^2},
\label{exoticsymp}
\end{equation}    
so that the Poisson bracket reads
\begin{equation}
    \big\{f, g\big\}=
    \Big(\frac{\partial f}{\partial x_{i}}\frac{\partial g}{\partial p_{i}}
    -
    \frac{\partial g}{\partial x_{i}}\frac{\partial f}{\partial p_{i}}\Big)
    +
    \theta\Big(
    \frac{\partial f}{\partial x_{1}}\frac{\partial g}{\partial x_{2}}
    -
    \frac{\partial g}{\partial x_{1}}\frac{\partial f}{\partial x_{2}}\Big).
\label{exoticPB}
\end{equation}
The spatial coordinates are therefore again non-commuting,
$
\big\{x, y\big\}=\theta
$
 cf. (\ref{xycomrel}).
 
Hamilton's equations associated with (\ref{exoticPB}) and with the 
standard free
Hamiltonian $H_{0}=\vp^2/2m$ describe the usual free motions; the 
``exotic'' structure only appears in the conserved quantities.
While the energy,
$H_{0}$, and the momentum, $\vp$, have the conventional form,
the angular momentum and Galilean boosts get indeed new contributions,
\begin{equation}
    \begin{array}{ll}
    &j=\vx\times\vp+\half\theta\,\vp^2,
    \\[4pt]
    &g_{i}=mx_{i}-p_{i}t+m\theta\,\epsilon_{ij}p_{j},
    \end{array}
\label{consquant}
\end{equation}

The commutation relations are those of the ``exotic''
[meaning two-fold centrally extended] Galilei group
which are the usual ones  except for the boosts which satisfy rather
\begin{equation}
    \big\{g_{1}, g_{2}\big\}=-m^2\theta=k.
\end{equation}

Conversely, positing the commutation relations 
(\ref{xycomrel}), augmented with the standard Heisenberg relations 
$\big\{x_{i}, p_{j}\big\}=\delta_{ij}$ and  $\big\{p_{i}, p_{j}\big\}=0$,
yields the unique symplectic form (\ref{exoticsymp}), showing that
exotic Galilean symmetry and non-commutative quantum mechanics are
indeed equivalent \cite{DH}.
     
Let us mention that the Hamiltonian structure presented here
for a free exotic particle is consistent with the ``exotic'' Lagrangian
\begin{equation}
    L_{0}=\vp\cdot\vx-\frac{\vp^2}{2m}+\frac{\theta}{2}\vp\times\dot{\vp}.
    \label{freeexoticlag}
\end{equation}    
    
The ``exotic'' extension plays hence little r\^ole for a free
particle, explaining (perhap) why it has only attracted little 
attention until recently. 
The situation changes dramatically, though, when
coupling to an abelian gauge field is considered.
Minimal coupling is achieved using Souriau's prescription \cite{SSD},
which simply means adding the electromagnetic $2$-form
 to the symplectic form of the system.
In terms of the Lagrangian, the minimally coupled expression reads
\begin{equation}
    L=\int(\vp-e\vA)\cdot\dot{\vx}-\frac{\vp^2}{2m}
    +eV+\frac{\theta}{2}\vp\times\dot{\vp}.
\label{mincouplag}
\end{equation}
 In the associated Euler-Lagrange equations,
\begin{equation}
    \begin{array}{ll}
    &m^*\dot{x}_{i}=p_{i}-em\theta\epsilon_{ij}E_{j},
    \\[4pt]
    &\dot{p}_{i}=eE_{i}+eB\epsilon_{ij}\dot{x}_{j},
    \end{array}
\label{mincoupleqmot}
\end{equation}
the extension parameters combine with the magnetic field
into and {\it effective mass},
\begin{equation}
    m^*=m\left(1-e\theta B\right).
    \label{effmass}
\end{equation}

For non-vanishing effective mass, $m^*\neq0$,
the motions are roughly similar to that of an ordinary particle
in a planar electromagnetic field \cite{DJT}.

When the effective mass, vanishes, $m^*=0$ i. e. for
\begin{equation}
    eB=\frac{1}{\theta},
    \label{criticalcase}
\end{equation}    
however, the system becomes singular, and the
consistency of the equations of motion (\ref{mincoupleqmot}) can 
only be maintained when
{\it all particles move according to the  Hall law}
\begin{equation}
    \dot{Q}_{i}=\varepsilon_{ij}\frac{E_{j}}{B}.
\label{Hallaw}
\end{equation} 
where $E_{i}=-\partial_{i}V$, and the $Q_{i}=x_{i}-E_{i}/B^2$,  
are suitable coordinates \cite{DH}.

Put in another way, symplectic\cite{DH,SSD}
(alias Hamiltonian\cite{FaJa}) reduction reduces
 the dimension of the phase space from $4$ to 
$2$, yielding namely the   
classical model (\ref{effdyn})
with $H$ given by the potential $eV$ and with
magnetic field $eB=\rho\delta\kappa$.

Geometrically, the $4$-dimensional phase space reduces to a 
two-dimensional one, namely to (\ref{hameq}).
Similarly, the minimally coupled Lagrangian (\ref{mincouplag})
reduces to the simple first-order Lagrangian (\ref{redlag}).

Note that our 
minimal coupling prescription is different from the one proposed
in Ref. 1 using the Poisson structure. 
This latter only allows a
constant magnetic field, and yields  conslusions
similar to but still different from ours.

\section{Quantization}

Then the Peierls substitution is recovered
by quantizing the reduced model \cite{DJT,DH},
 conveniently carried out in the Bargmann-Fock
framework. Setting $z=\theta^{-1/2}(x+iy)$, and
choosing the holomorphic polarization, the
the wave functions are $f(z)e^{-\vert z\vert^2/4}$ with $f(z)$
analytic; the fundamental (creation and annihilation) operators 
$ 
\widehat{z}=z\cdot,
\, 
\widehat{\bar{z}}=2\partial_{z}
$ 
satisfy $[\widehat{\bar{z}},\widehat{z}]=2$.

Finding the spectrum requires quantizing  the Hamiltonian
 $H=eV$. 
The point is that  the answer depends crucially on the chosen
quantization scheme.
Let us restrict ourselves to the radial case $V=U(r^2)$.
Then the classical system is symmetric w. r. t. rotations.
The conserved angular momentum for the planar system
(\ref{hameq}) is in particular
\begin{equation}
J=\frac{1}{2\theta} r^2=\half z\bar{z}.
\label{angmom}
\end{equation}
The system has therefore fractional angular momentum,   
$j_{n}=\alpha_{0}+n,\, n=0, 1,\dots$, as it is readily derived\cite{HMC}
using the representation theory of the symplectic group sp$(1)$.

Let us now turn to the quantization of the radial potential 
$U(r^2)=U(2\theta J)$.
One of the schemes\cite{NC2,HMC} says that the spectrum of is simply
$ 
U(2\theta j_{n}).
$ 
This scheme ignores, however, the problem of operator ordering.
According to another proposal\cite{GJ,DJT}, $\widehat{H}$ is obtained
by {\it anti-normal ordering}. In terms of the complex coordinates 
$z$ and $\bar{z}$, this amounts to ``putting all the
$\widehat{\bar{z}}$ to the left and all the $\widehat{z}$ to the right''.

Recently\cite{DH} we argued that 
this prescription requires further modification,
and proposed to use instead 
Bergman  quantization\cite{Bergman}, which identifies
the quantum operator associated to $H(z,\bar{z})$ as
\begin{equation}
    \widehat{H}\psi(z)=
    \int\exp\left[\half\bar{\zeta}(z-\zeta)\right]
    \left(H-\partial_{z}\partial_{\bar{z}}H\right)
    \psi(\zeta)d{\zeta}d\bar{\zeta}.
    \label{Bergman}
\end{equation}

Although we have not yet evaluated this for
the vortex Hamiltonian (\ref{vortham}),
it is obvious that it will be different from the
simple formula above.
When $V$ is a polynomial, the prescription (\ref{Bergman})
simplifies to first subtracting a correction term,
$V\to\widetilde{V}=V-\partial_{z}\partial_{\bar{z}}V$,
and then anti-normal order $\widetilde{V}$ \cite{Bergman}.

The model discussed here has been used in the context of
the Fractional Quantum Hall Effect, namely 
to calculate the ``1/3 effect''\cite{GJ,QHE}
using anti-normal ordering. 
There the LLL electrons were supposed to be in pairwise 
Coulomb interaction. For two quasiparticles, this means $V=r^{-1}$. 
Bergmann quantization yields\cite{Bergman} instead that the operator 
$\widehat{V}$  is multiplication with
\begin{equation}
\frac{1}{r}\,{\rm erf}(\sqrt{r/\theta})+(\theta\pi)^{-1/2}
e^{-r^2/\theta},
\label{BCoulop}
\end{equation}
which only differs from the usual expression in the region 
$r<\theta$. The consequences of this short-distance modifications
have not yet been drawn.

\vskip2mm
\noindent{\bf Acknowledgments}.
Most of the results presented here stem from
joint work with C. Duval, Z. Horv\'ath, L. Martina, and M. Plyushchay.
It is a pleasure to acknowledge the organizers of
International Workshop {\it Nonlinear Physics: Theory and Experiment. 
}{\rm II}. Gallipoli, (Lecce, Italy)
for their kind hospitality extended to me during the Workshop.

\goodbreak

\end{document}